\documentclass[sigconf]{acmart}

\usepackage[title]{appendix}
\UseRawInputEncoding

\AtBeginDocument{%
  \providecommand\BibTeX{{%
    \normalfont B\kern-0.5em{\scshape i\kern-0.25em b}\kern-0.8em\TeX}}}

\copyrightyear{2020}
\acmYear{2020}
\setcopyright{none}
\acmConference[FDG '20]{International Conference on the Foundations of Digital Games}{September 15--18, 2020}{Bugibba, Malta}
\acmBooktitle{International Conference on the Foundations of Digital Games (FDG '20), September 15--18, 2020, Bugibba, Malta}
\acmPrice{15.00}
\acmDOI{10.1145/xxxxxxx.xxxxxxx}
\acmISBN{978-1-4503-8807-8/20/09}

\begin{document}

\title{Hack.VR: A Programming Game in Virtual Reality}

\author{Dominic Kao*, Christos Mousas*, Alejandra J. Magana*, D. Fox Harrell$\dagger$, Rabindra Ratan$\ddagger$, \,\,\,\, Edward F. Melcer$\mathsection$, Brett Sherrick*, Paul Parsons*, Dmitri A. Gusev*}
\affiliation{%
  \institution{* Purdue University, $\dagger$ MIT, $\ddagger$ Michigan State University, $\mathsection$ UC Santa Cruz}
  \city{kaod@purdue.edu, cmousas@purdue.edu, admagana@purdue.edu, fox.harrell@mit.edu, rar@msu.edu, eddie.melcer@ucsc.edu, bsherrick@purdue.edu, parsonsp@purdue.edu, dgusev@purdue.edu}
}

\renewcommand{\shortauthors}{Kao et al.}

\begin{abstract}
In this article we describe Hack.VR (\textquotesingle hæk \textquotesingle d\raisebox{-0.25px}{\includegraphics[height=4.25px]{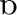}}t /vr/), an object-oriented programming game in virtual reality. Hack.VR uses a VR programming language in which nodes represent functions and node connections represent data flow. Using this programming framework, players reprogram VR objects such as elevators, robots, and switches. Hack.VR has been designed to be highly interactable both physically and semantically.

\end{abstract}



\begin{teaserfigure}
  \includegraphics[width=\textwidth]{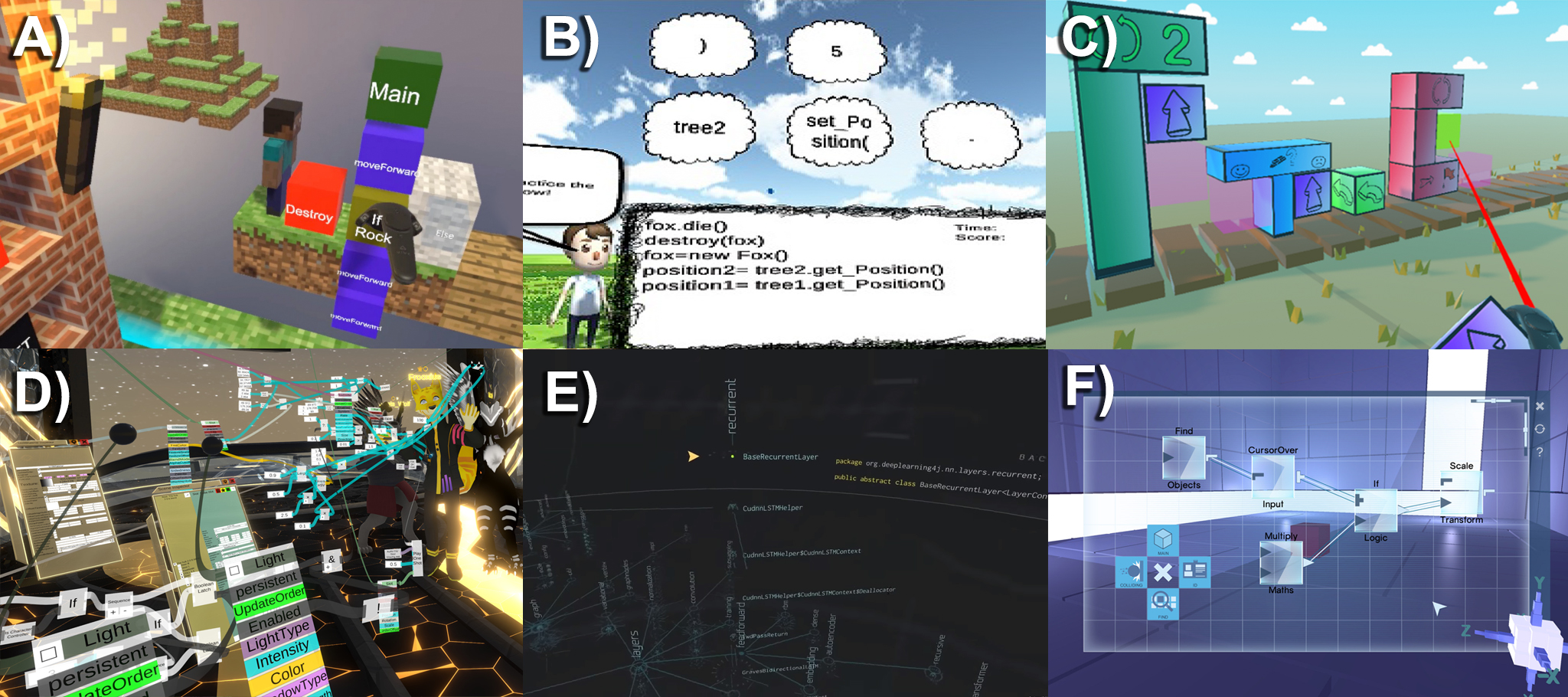}
\vspace{-18px}
  \caption{Existing environments for programming in VR: A) Block-based programming (Cubely), B) Multiple-choice code snippet selection (Imikode), C) Block-based programming (VR-OCKS), D) World-building (NeosVR), E) Code visualization (Primitive), F) Visual programming (Glitchspace). All figures reproduced with permission.}
  \label{fig:VRSystems}
\vspace{8px}
\end{teaserfigure}

\maketitle

\begin{figure*}
  \includegraphics[width=\textwidth]{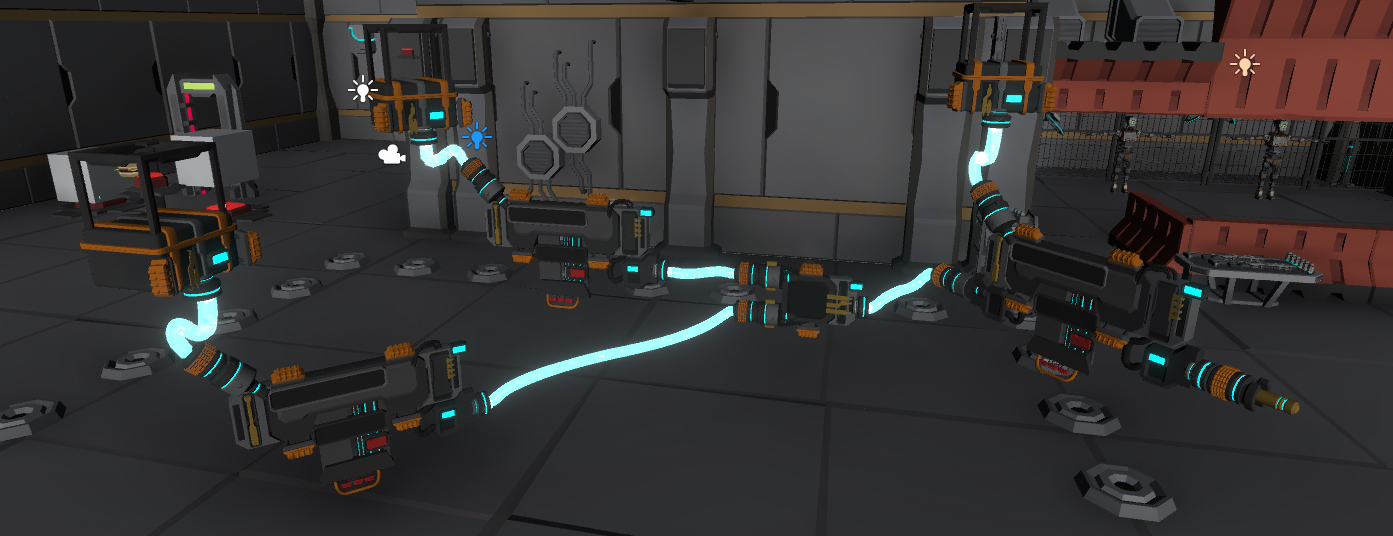}
  \vspace{-18px}
  \caption{In Hack.VR, programs are sets of nodes. Nodes are interconnected by blue tubes, which represent data flow.}
  \label{fig:ProgramsSetsOfNodes}
\end{figure*}

\vspace{-10px}

\section{Introduction}
In the entire history of computing, programming has been a largely physically static activity. But technologies previously inaccessible to most users are now growing rapidly. Today, 78\% of Americans are familiar with VR (from 45\% in 2015) \cite{GreenlightInsights}. As a result,  experiences traditionally created for desktops are now appearing in VR, e.g., training \cite{Bertram2015} and automobile design \cite{Lawson2016}. Researchers argue that VR increases immersion \cite{hussein2015benefits}, which in turn increases engagement and learning \cite{Dede2009}. VR might be especially useful for teaching programming because spatial navigation in VR helps reduce extraneous cognitive load and increase germane cognitive focus on learning content compared to text on a screen \cite{Lee2014}. Further, VR allows users to experience a sense of self-presence in the environment \cite{ratan2013self}, which facilitates an embodied-cognitive learning experience \cite{Shin2017,melcer2018learning} through which users interact with the learning content more intuitively \cite{Steed2016a}, potentially augmenting learning outcomes \cite{cheon2012effects}. Nonetheless, only a handful of environments for programming exists in VR. In this article, we describe a programming game in virtual reality that we created called Hack.VR.\footnote{Trailer video: \url{https://youtu.be/3Mp6ISjD1mg}.}\footnote{Walkthrough video: \url{https://youtu.be/TGc8H3Nw-3M}.}

\section{Programming in VR}

Existing environments for programming in VR can be seen in Figure~\ref{fig:VRSystems}. These include VR-OCKS and Cubely, block-based VR programming environments \cite{Segura2019,Vincur2017}, and Imikode, a multiple choice code snippet selection environment in VR \cite{Oyelere2019}. Other significant projects include NeosVR \cite{NeosVR2020}, a shared social universe that features powerful programming tools for VR world creation, and Primitive \cite{Primitive2020}, an ``immersive development environment'' enabling 3D code visualization. In the indie game Glitchspace \cite{Glitchspace2016}, players use a visual programming language to solve puzzles. These environments for programming in VR have been developed for education (Cubely, Imikode, VR-OCKS), for modifying virtual worlds (NeosVR), for code visualization (Primitive), and for entertainment (Glitchspace).

Importantly, Hack.VR was created specifically to teach \textit{object-oriented programming} (OOP) compared to the highly procedural approaches in the systems above. OOP encapsulates logic into objects. This paradigm has a natural translation to VR, where 3D objects can each contain internal programming that is abstracted from observers. Hack.VR is the first system for learning OOP in VR, and will serve as a testbed to perform research studies. This testbed may also be useful for studying other aspects, e.g., help facilities, embellishment, the player avatar, feedback, and the resulting effects on VR programming \cite{Frommel2017,Kao2020a,Hicks2019a,Kao2020,Kao2019c,Kao2017,Birk2016a,Kao2017b,Kao2019a,Kao2018,Kao2016e,Kao2015,Kao2015a,ORourke2014,kaoexploring,Kao2016f}.

\section{The Game}

\subsection{Engine}

\begin{figure}
  \includegraphics[height=0.36\columnwidth]{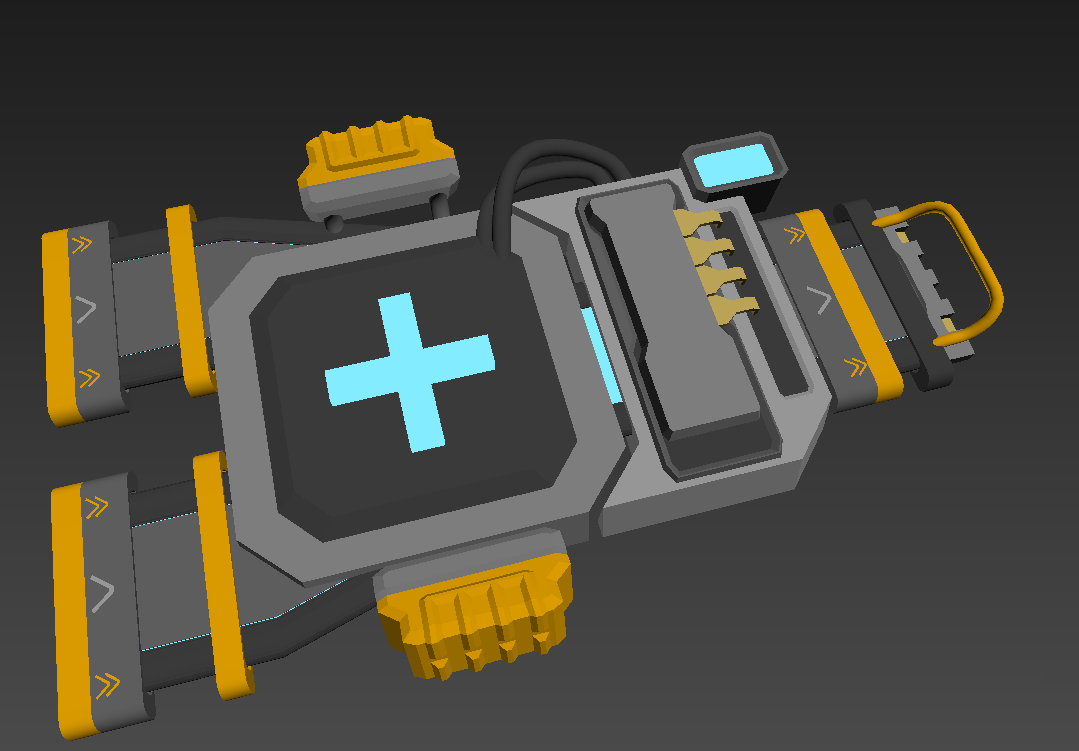}\includegraphics[height=0.36\columnwidth]{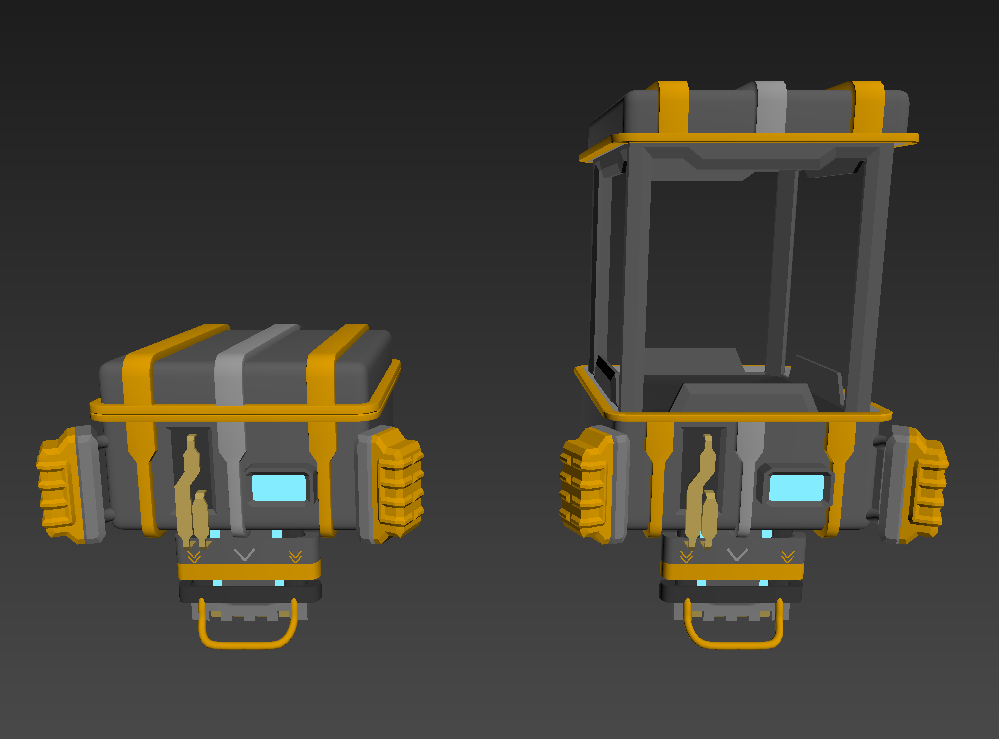}
    \vspace{-16px}
  \caption{Left: An arithmetic node that performs addition. Right: An entity node contains a miniaturized representation of a virtual world object, e.g., a door, a robot, an elevator. Programs attached to an entity node will then operate on the actual virtual world object, e.g., open the door, turn the robot, operate the elevator.}
  \label{fig:ExampleNodes}
\end{figure}

\begin{figure}
\offinterlineskip
  \includegraphics[width=\columnwidth]{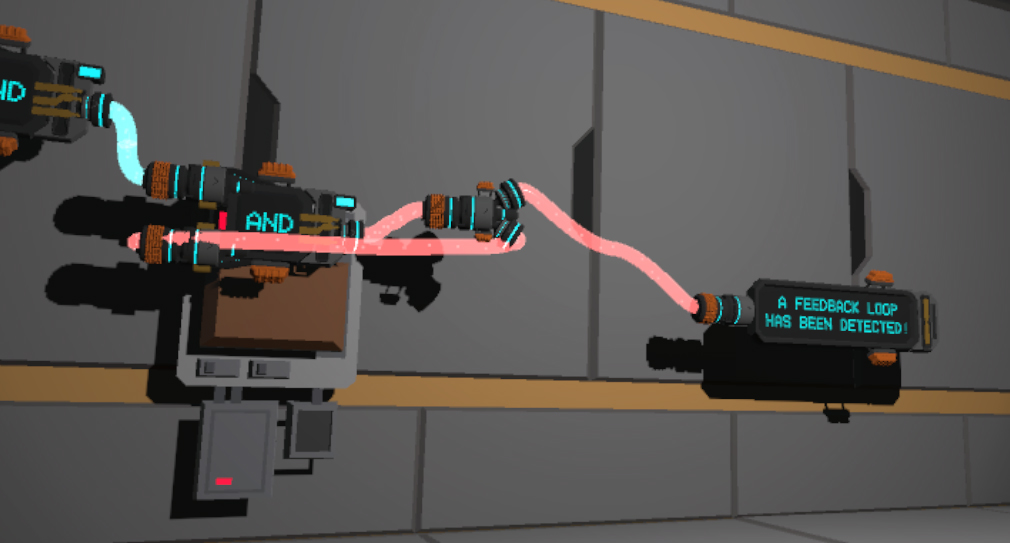}
  \includegraphics[width=\columnwidth]{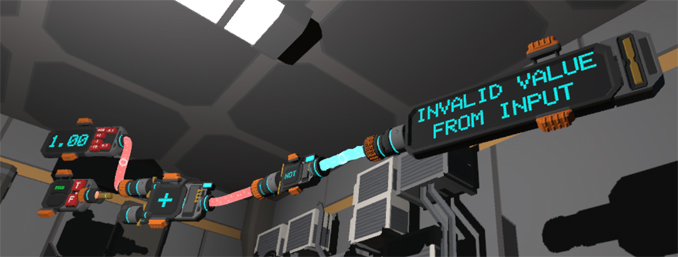}
  \includegraphics[width=\columnwidth]{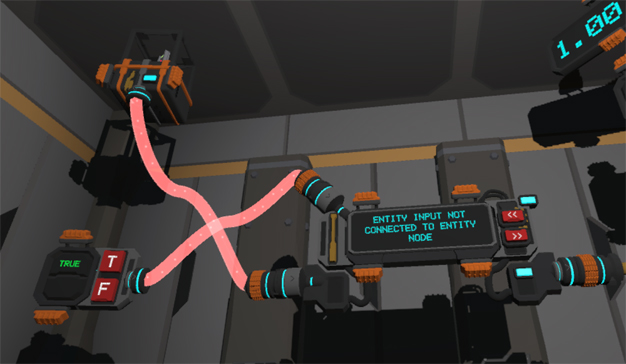}
  \vspace{-6px}
  \caption{Example program errors. Top: A feedback loop. Middle: An invalid input. Bottom: Another invalid input. Errors are automatically displayed in the node after the error.}
  \label{fig:Errors}
\end{figure}

In Hack.VR, a program is a set of nodes. See Figure~\ref{fig:ProgramsSetsOfNodes}. Nodes contain typical programming constructions, e.g., primitive values, objects, arithmetic operators, conditionals, event handlers, function calls. Nodes facilitate communication through data flow. Nodes may have both inputs and outputs depending on the node type. For example, see Figure~\ref{fig:ExampleNodes}. Nodes can also represent entire method calls, the details of which are abstracted from the player except input and output. Because the goal of Hack.VR is to teach the player OOP, the inner-workings of the methods themselves are intentionally abstracted away (and players cannot see the code) so that the player can concentrate on higher-level representations. The engine also supports extensions. For example, once a new function has been defined in the engine, a node can call it. To reduce complexity, players in Hack.VR use designer-embedded nodes to solve each puzzle instead of creating their own nodes. While the engine supports on-the-fly node creation, the UI does not currently support this. Node-based programming, like any other type of programming, can lead to execution errors. For example, a \textit{NOT} node expects a boolean input (true or false), and outputs the inversion of its input. However, if a numerical value is instead used as input to a \textit{NOT} node, this results in an error. While this is valid in some procedural languages (in C programming, 0 is false and anything else is true), implicit conversions from numerical values to boolean data types is not allowed in object-oriented programming (e.g., Java, C\#). When an error is detected in the program, this is indicated by a red tube. See Figure~\ref{fig:Errors} for examples.

\begin{figure}
\includegraphics[height=0.53\columnwidth]{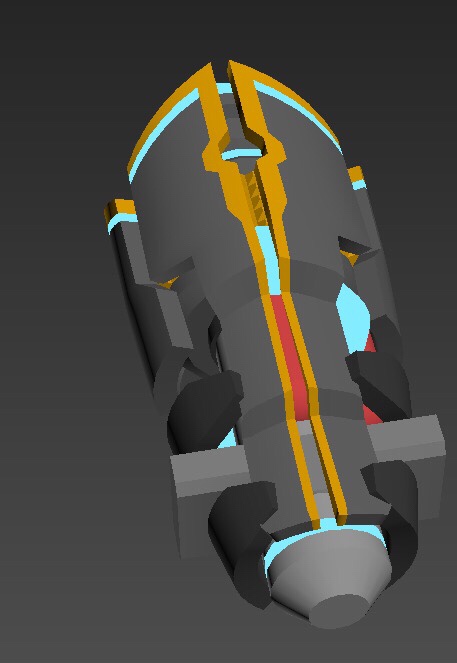}\includegraphics[height=0.53\columnwidth]{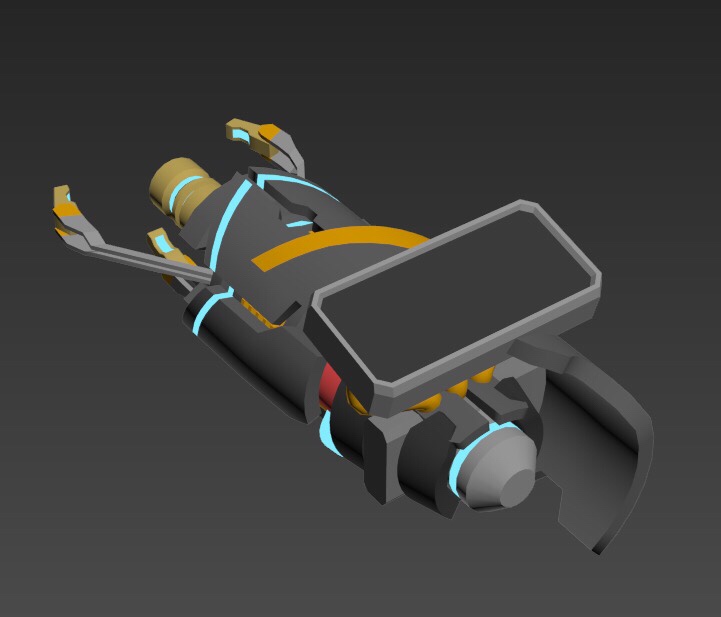}
  \caption{Left: The left gun allows the player to freely teleport. Right: The right gun allows the player to manipulate objects, modify programs, and inspect nodes. Inspecting shows a node's output value on the square display.}
  \label{fig:Guns}
\end{figure}

\begin{figure}
\includegraphics[width=0.5\columnwidth]{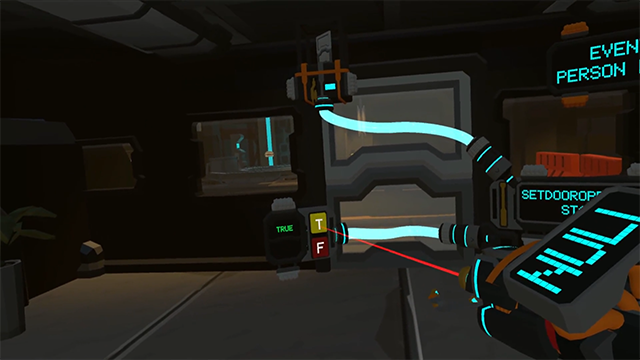}\includegraphics[width=0.5\columnwidth]{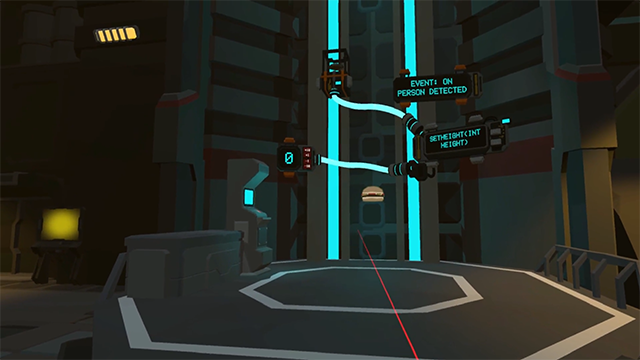}
  \vspace{-20px}
  \caption{Left: Puzzle 1 is opening a door using a boolean. Right: Puzzle 2 is setting the height of an elevator so you can reach the next puzzle.}
  \label{fig:Puzzles1}
    \vspace{5px}
  \includegraphics[width=0.5\columnwidth]{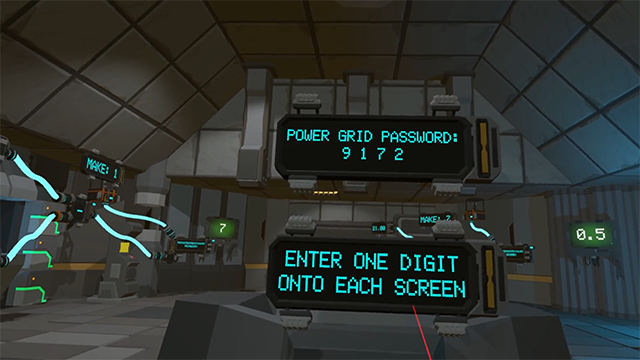}\includegraphics[width=0.5\columnwidth]{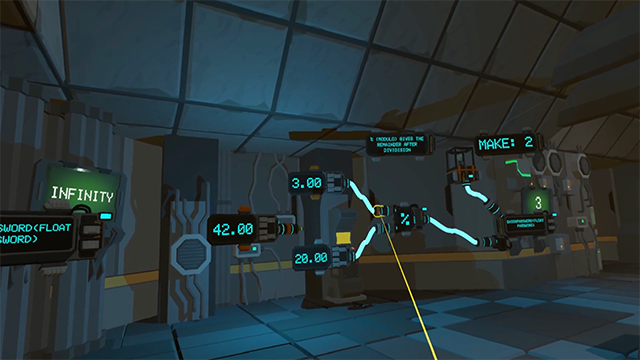}
    \vspace{-20px}
  \caption{Puzzle 3 is four mini-puzzles where the player manipulates arithmetic nodes to create a password.}
      \vspace{5px}
  \includegraphics[width=0.5\columnwidth]{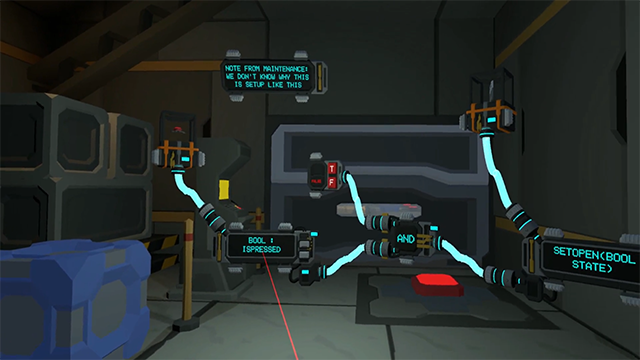}\includegraphics[width=0.5\columnwidth]{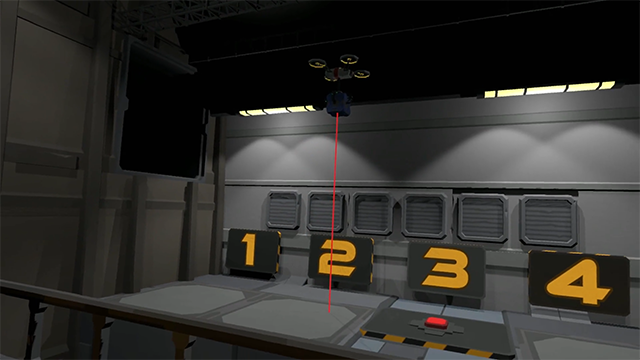}
  \vspace{-20px}
  \caption{Puzzles 4--8  teach logical operators. Left: A puzzle using the \texttt{and} operator. Right: Programming a robot to drop a cube only on the column marked 3.}
      \vspace{5px}
  \includegraphics[width=0.5\columnwidth]{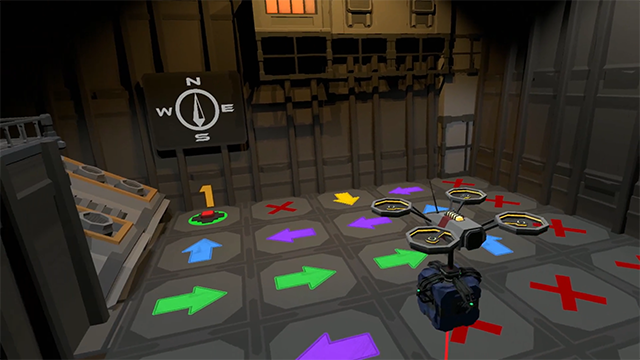}\includegraphics[width=0.5\columnwidth]{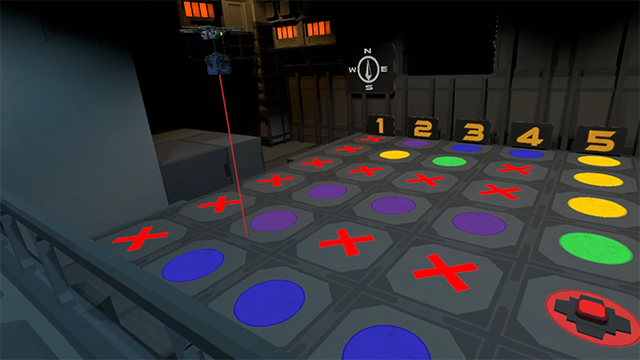}
    \vspace{-20px}
  \caption{Left: Puzzle 9 requires programming a robot to move based on color. Arrows indicate in which direction the robot should move. Right: Puzzle 10 is an extension of Puzzle 9, and now  the robot moves based on a combination of column number and color.}
        \vspace{5px}
  \includegraphics[width=0.5\columnwidth]{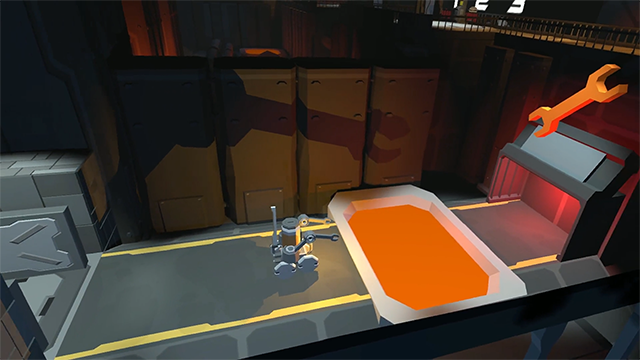}\includegraphics[width=0.5\columnwidth]{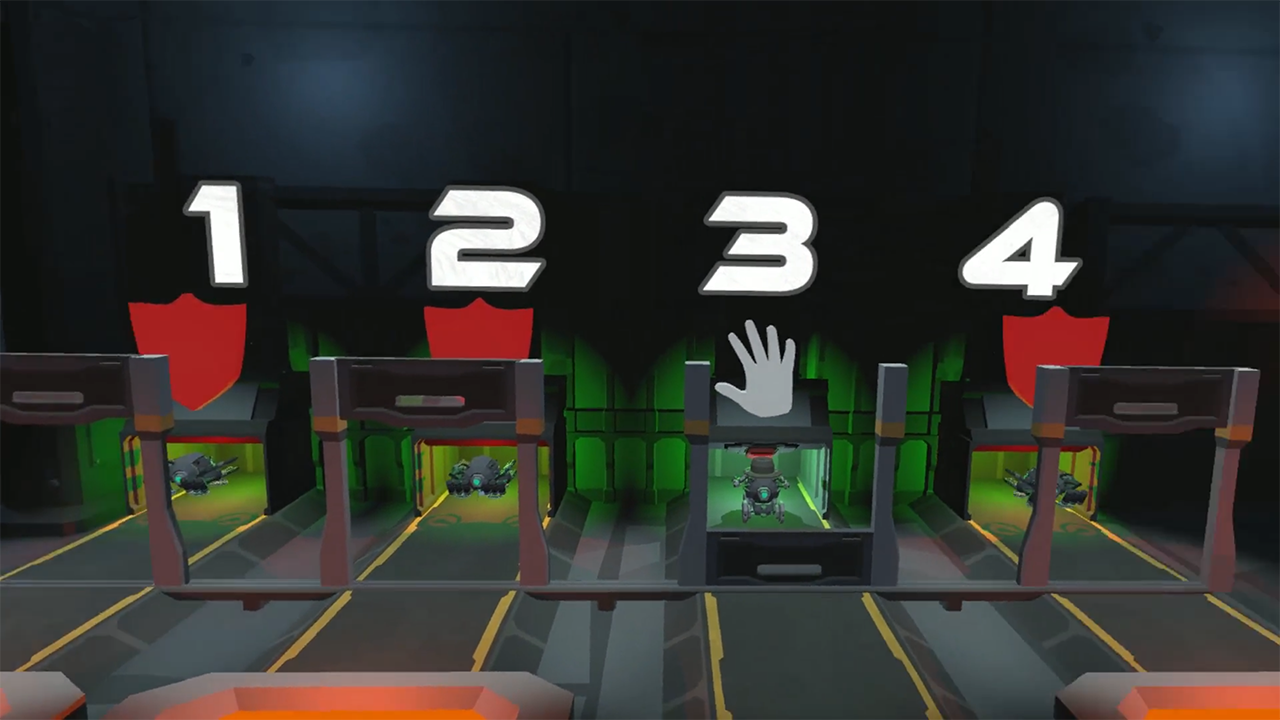}
    \vspace{-20px}
  \caption{Puzzles 11, 12, and 13 explore constructors. Left: The robot in Puzzle 11 must be constructed with the \texttt{hover} movement type to get past the lava. Right: Robots in Puzzle 13 must be constructed with correct movement types and body types matching each lane.}
\end{figure}

Similarities can be drawn between Hack.VR and other programming paradigms inherently predisposed to visualization, e.g., flow-based programming \cite{morrison1994flow}. Hack.VR is inspired by node graph systems: Unreal Engine 4's Blueprints \cite{EpicGames2020}, Unity Playmaker \cite{HutongGamesLLC2019}, and shader editors (Shader Graph \cite{UnityTechnologies2020} and Amplify Shader Editor \cite{AmplifyCreations2020}).

\begin{figure}
\setlength{\lineskip}{0pt}
  \includegraphics[height=0.2618\columnwidth]{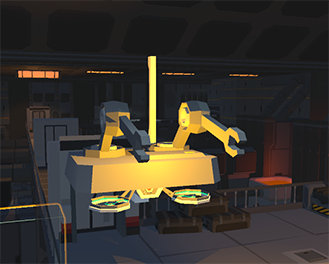}\includegraphics[height=0.2618\columnwidth]{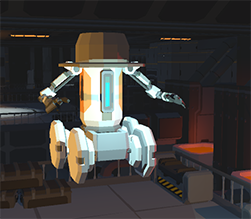}\includegraphics[height=0.2618\columnwidth]{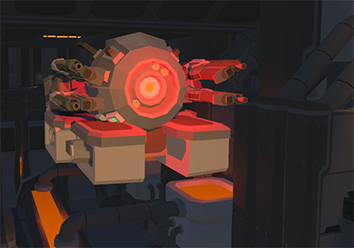}
   \includegraphics[width=\columnwidth]{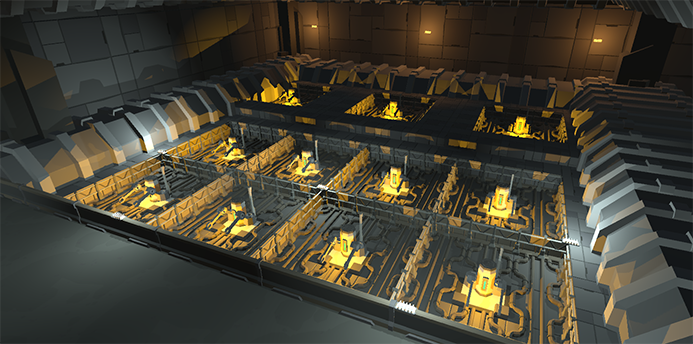}
    \vspace{-20px}
  \caption{Puzzles 14--17 explore classes and inheritance. In Puzzle 14, the player modifies a class to modify all objects that are created from that class.}

  \label{fig:Puzzles2}
\end{figure}

\subsection{Art Style and Gameplay}

\begin{figure}
\setlength{\lineskip}{0pt}
\includegraphics[width=\columnwidth]{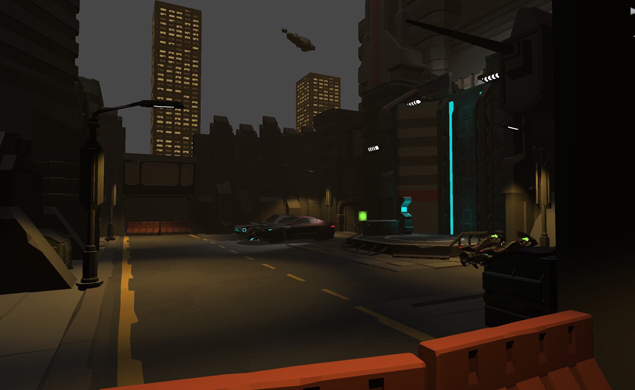}
\includegraphics[width=\columnwidth]{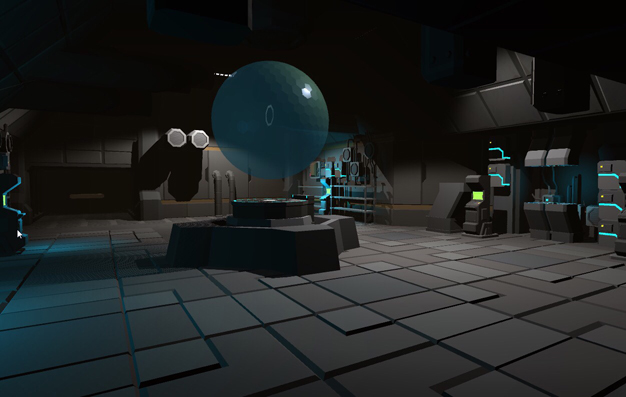}
    \vspace{-18px}
  \caption{Hack.VR's sci-fi futuristic setting.}
  \label{fig:Environments}
\end{figure}

Hack.VR is based in a sci-fi futuristic setting. See Figure~\ref{fig:Environments}. In Hack.VR, the player holds two ``guns'' that take the place of their VR controllers. See Figure~\ref{fig:Guns}. Hack.VR is compatible with both HTC Vive and Oculus. Hack.VR's controls are found in Appendix A. Using these controls, Hack.VR challenges players with object-oriented programming puzzles. Hack.VR consists of 17 different puzzles. Each puzzle builds upon concepts from prior puzzles. See Figures~\ref{fig:Puzzles1} through \ref{fig:Puzzles2} for a short description of puzzles. %

\subsection{Design Process}

\begin{figure}
\includegraphics[width=\columnwidth]{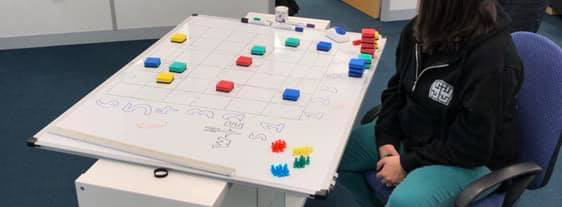}
    \vspace{-18px}
  \caption{Prototyping Puzzles 9 and 10.}
  \label{fig:Design}
\end{figure}

The design process followed a spiral approach of design \textrightarrow~implementation \textrightarrow~evaluation. Iterations grew in complexity and refinement over several cycles for each part of the game. Feedback was solicited from designers, developers, and playtesters. Comments affirmed positive design choices (e.g., ``I like that you can see the physical buttons and physical door miniaturized in the node tree'') and highlighted potential improvements (e.g., ``When I'm connecting things, it's hard to tell what connector I'm working with; maybe highlight the current selected connector?''). A typical early prototype can be seen in Figure~\ref{fig:Design}.

\section{Conclusion}

In this article we described Hack.VR, a programming game in virtual reality. We created a VR programming language that is highly visual, while being highly semantic. Hack.VR is inspired by the possibilities of programming in VR. Imagine these highly evocative scenarios:
\begin{itemize}
   \item Programming an infinite stairwell taking you into the clouds.
   \item Programming a robot carrying you across vast deserts, rolling hills, and tundras.
   \item Reconfiguring and reprogramming the mechanical parts in your gun to enhance your capabilities.
\end{itemize}

Given the great potential for VR to enhance learning outcomes \cite{Lee2014,Shin2017,Steed2016a,cheon2012effects}, we expect that Hack.VR might help teach programming concepts more effectively than similar, non-immersive tools. Although assessment research should be conducted to confirm this expectation empirically, from a perspective that spans research, design, and play, there is reason to be excited about what the coming decade will bring for programming in VR.

\hphantom{T}
\hphantom{T}
\hphantom{T}
\hphantom{T}
\hphantom{T}

\bibliographystyle{ACM-Reference-Format}
\bibliography{bib}

\onecolumn

\section*{APPENDIX A}

\subsection*{Hack.VR Controls}

\newcommand{\mysize}{0.155} %
\captionsetup{justification=centering,margin=2.5cm}

\begin{figure}[H]
\includegraphics[height=\mysize\columnwidth]{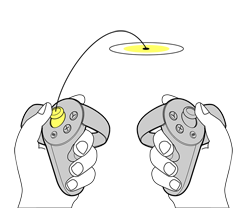}\includegraphics[height=\mysize\columnwidth]{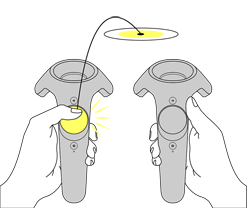}
    \vspace{-4px}
  \caption{Movement.}
  \label{fig:Controls1}
\end{figure}

\vspace{-10px}

\begin{figure}[H]
\includegraphics[height=\mysize\columnwidth]{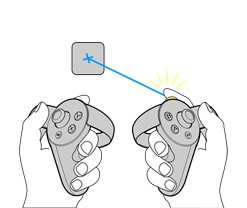}\includegraphics[height=\mysize\columnwidth]{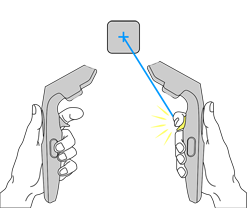}
    \vspace{-4px}
  \caption{Grabbing an object. Once an object is grabbed, it can be rotated and moved. Grabbing a  node will inspect it, displaying its output on the right gun.}
\end{figure}

\vspace{-10px}

\begin{figure}[H]
  \includegraphics[height=0.120\columnwidth]{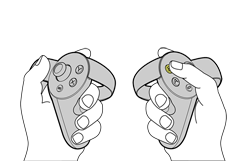}\includegraphics[height=0.120\columnwidth]{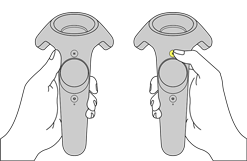}
    \vspace{-4px}
  \caption{Switch between physical manipulation and program modification mode. A blue laser indicates physical manipulation, while a red laser indicates modification.}
\end{figure}

\vspace{-10px}

\begin{figure}[H]
  \includegraphics[height=\mysize\columnwidth]{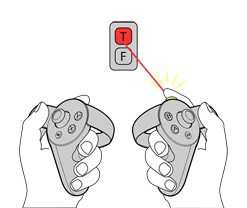}\includegraphics[height=\mysize\columnwidth]{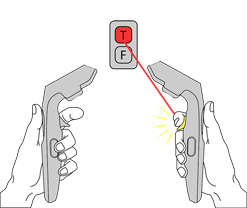}
    \vspace{-4px}
  \caption{Program modification.}
\end{figure}

\vspace{-10px}

\begin{figure}[H]
  \includegraphics[height=0.163\columnwidth]{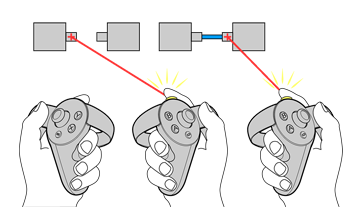}\includegraphics[height=0.163\columnwidth]{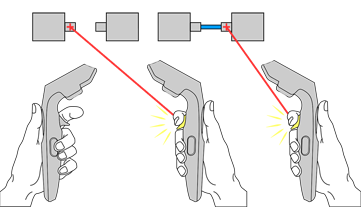} %
    \vspace{-4px}
  \caption{Modifying node attachments.}
  \label{fig:Controls2}
\end{figure}

\end{document}